# Localized Thermal Gradients On-Chip by Radiative Cooling of Silicon Nitride Nanomechanical Resonators

Alexandre Bouchard[a], Timothy Hodges[a,b], Michel Stephan[a], Lixue Wu[b], Triantafillos Koukoulas[b], Richard Green[b], Raphael St-Gelais[a,c,d]

[a] Department of Mechanical Engineering, University of Ottawa, 75 Laurier Avenue East, Ottawa, Ontario K1N 6N5, Canada
[b] Metrology Research Centre, National Research Council Canada, 1200 Montreal Road, Ottawa, Ontario K1A 0R6, Canada
[c] Centre for Research in Photonics, University of Ottawa, 75 Laurier Avenue East, Ottawa, Ontario K1N 6N5, Canada
[d] Department of Physics, University of Ottawa, 75 Laurier Avenue East, Ottawa, Ontario K1N 6N5, Canada

**Abstract**

Small scale renewable energy harvesting is an attractive solution to the growing need for power in remote technological applications. For this purpose, localized thermal gradients on-chip—created via radiative cooling—could be exploited to create microscale renewable heat engines running on environmental heat. This could allow self-powering in small scale portable applications, thus reducing the need for non-renewable sources of electricity and hazardous batteries. In this work, we demonstrate the creation of a local thermal gradient on-chip by radiative cooling of a 90 nm thick freestanding silicon nitride nanomechanical resonator integrated on a silicon substrate that remains at ambient temperature. The reduction in temperature of the thin film is inferred by tracking its mechanical resonance frequency, under high vacuum, using an optical fiber interferometer. Experiments were conducted on 15 different days during fall and summer months, resulting in successful radiative cooling of the membrane in each case. Maximum temperature drops of 9.3 K and 7.1 K are demonstrated during the day and night, respectively, in close correspondence with our heat transfer model. Future improvements to the experimental setup could improve the temperature reduction to 48 K for the same membrane, while emissivity engineering potentially yields a maximum theoretical cooling of 67 K with an ideal emitter.

## 1    Introduction

Converting ambient heat to electricity on-chip is an attractive solution to the increasing need for small scale remote power in delocalized applications. However, from a thermodynamic standpoint, harvesting ambient heat is only possible with a heat sink at a colder temperature. Outer space ($T \approx 3$ K) is a sustainable prospect for such a heat sink. Remarkably, objects on Earth can access this heat sink via radiative cooling, which is possible due to atmospheric transparency ($\tau_{\lambda,atm}$) in the 8 μm – 13 μm wavelength window (see Fig. 1a).

Radiative cooling has been investigated in applications such as heat extraction (e.g., in buildings [1]–[6] and photovoltaics [7]–[10]) and macroscale energy harvesting [11]–[14], but its use for microscale electricity generation integrated on-chip has not been reported. In the 1970's and 1980's, radiative cooling of bulk materials with intrinsic radiative cooling properties was studied for its potential energy savings applications [15]–[21]. The cooling of such bulk materials was mostly limited to nighttime radiative cooling, or in some cases daytime cooling with diffuse solar irradiation only. More recently, advances in material science and photonic devices have allowed for more tuning of the spectral radiative properties of an object. This has made daytime radiative cooling with direct sunlight exposure possible by suppressing absorption in the solar spectrum [7], [22]–[35]. Raman et al. first demonstrated radiative cooling under direct sunlight with a 1D multilayer structure in 2014, achieving a temperature drop of 4.9 K below ambient temperature [23]. In 2016, Chen et al. demonstrated a record temperature drop of 42 K below ambient in peak daylight [25]. Energy harvesting applications harnessing ambient heat via radiative cooling have been demonstrated, but only by cooling of a macroscopic object connected to an external, bulk, thermoelectric generator [12], [13]. Full integration of such a heat engine on a chip would require localized cooling—i.e., a thermal gradient on-chip—with integrated thermoelectric circuits. This was not possible in the previously investigated devices where the entire substrates were uniformly cooled.

Here we demonstrate the first fundamental ingredient of a microscale integrated heat engine—creation of a local thermal gradient on-chip—by radiative cooling of a 90 nm thick freestanding silicon nitride (SiN) membrane integrated on a silicon (Si) chip that remains at ambient temperature (see Fig. 1b). Recent work has shown that the thermal coupling of a SiN membrane with its environment can be dominated by thermal radiation [36]–[38], making SiN a material platform of choice for thermal radiation sensing experiments [39], [40]. Moreover, the radiative hemispherical emissivity spectrum of freestanding SiN ($\varepsilon_{\lambda,m}$) is naturally well aligned with the atmospheric transparency window—in the 8 μm – 13 μm wavelength range—while having weak absorptivity elsewhere (see Fig. 1a). As such, freestanding SiN membranes possess the criteria for significant radiative cooling: (i) high emissivity in the 8 μm – 13 μm wavelength range to maximize thermal radiation to outer space through the atmospheric transparency window, (ii) weak absorptivity outside of the 8 μm – 13 μm wavelength range to minimize parasitic radiative heating (e.g., from the atmosphere, the sun, etc.) [3], and (iii) radiation-dominated thermal coupling with the environment to minimize conductive heating. Moreover, SiN membranes are a robust platform over which it is possible to deposit metal and semiconductors to create planar thermoelectric circuits [41], which could enable micro heat engines on-chip.

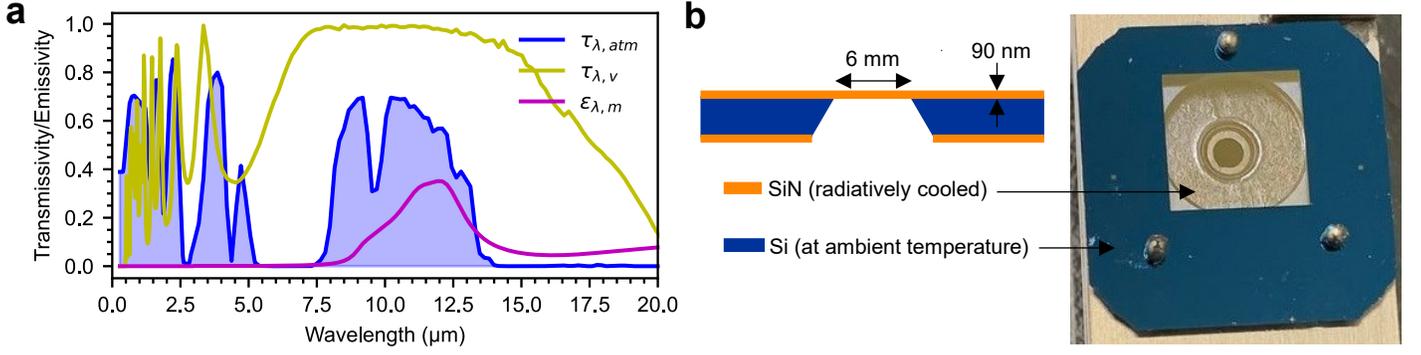

Fig. 1. a) $\tau_{\lambda,atm}$: atmospheric transmissivity at normal incidence for mid-latitude summer climate conditions computed by the Moderate Resolution Atmospheric Transmission (MODTRAN) model [42], [43]. $\tau_{\lambda,v}$: zinc selenide (ZnSe) viewport spectral transmissivity at normal incidence (Thorlabs WG71050-E3 [44]). $\tau_{\lambda,v}$ and $\tau_{\lambda,atm}$ are assumed null at wavelengths above 20 μm. $\varepsilon_{\lambda,m}$: spectral hemispherical emissivity of a 90 nm thick SiN membrane (as shown in b) calculated using the procedure described in section 2.1. b) Structurally released 6 mm × 6 mm, 90 nm thick, freestanding SiN nanomechanical resonator used in this work, where the SiN thin film is radiatively cooled and the silicon substrate remains at ambient temperature.

## 2    Heat Transfer Model

A depiction of the radiative cooling experiment conducted in this work—a freestanding SiN membrane mounted inside a vacuum chamber and facing the sky through a zinc selenide (ZnSe) viewport—is presented in Fig. 2a. The main heat transfer rates are also depicted in Fig. 2a, where $q_m$ is the rate at which radiation is leaving the membrane, $q_{atm}$ is the rate at which atmospheric irradiation is incident on the membrane, $q_{ch,top}$ and $q_{ch,back}$ are the rates at which chamber irradiation from the topside and backside of the membrane, respectively, are incident on the membrane, $q_{cond}$ is the rate of conduction from the silicon frame to the membrane, and $q_{sun}$ is the rate at which solar irradiation is incident on the membrane. Convection is negligible because the membrane is mounted inside a vacuum chamber which is kept at high vacuum in the $10^{-6}$ hPa range (see Supplementary Note 1). The net cooling rate (in W/m²) of the membrane is then given by:

$$q_{net}(T_m, T_{atm}, T_{ch}, \theta_v, \theta_{sun}) = q_m - q_{atm} - q_{ch,top} - q_{ch,back} - q_{cond} - q_{sun}, \qquad (1)$$

where $T_m$, $T_{atm}$, and $T_{ch}$ are the respective temperatures of the membrane, atmosphere, and chamber, $\theta_v$ is the viewport angle representing the field of view of the membrane (see Fig. 2a), and $\theta_{sun}$ is the angle at which the sun is facing the membrane (see Fig. 2a). For given values of $T_{atm}$, $T_{ch}$, $\theta_v$, and $\theta_{sun}$, the steady state energy balance (i.e., $q_{net} = 0$) can be solved to determine the theoretical membrane temperature ($T_m$)—and thus the temperature drop ($\Delta T_m = T_m - T_{ch}$)—from radiative cooling. For this purpose, the expressions for all heat transfer rates are detailed in section 2.2.

### 2.1    Membrane Emissivity Models

The membrane spectral directional emissivity ($\varepsilon_{\lambda,\theta,m}$) is a central quantity for evaluating the net radiative cooling rate in eq. (1), and its value can be affected by the membrane surroundings. Specifically, the backside surrounding of the membrane (i.e., the optical fiber holder and membrane mount; see Fig. 2a) is an unpolished, partially reflective, metal surface that can act both as (i) a mirror that enhances coupling of the membrane with the sky and improves cooling or, (ii) a parasitic source of radiated heat. Precisely quantifying these competing effects is not immediately obvious, so we account for the two extreme cases. In a best-case scenario, the backside is an aluminum mirror that increases the effective emissivity of the membrane (see Fig. 2b). In a worst-case scenario, the membrane is a large isothermal surrounding emitting blackbody radiation (see Fig. 2b). These cases result in different membrane emissivities, which ultimately affect the net cooling rate of the membrane. In both cases, the membrane spectral directional emissivity is calculated using Kirchhoff's law and surface radiation balance as [45]:

$$\varepsilon_{\lambda,\theta,m} = \alpha_{\lambda,\theta,m} = 1 - \tau_{\lambda,\theta,m} - \rho_{\lambda,\theta,m}, \qquad (2)$$

where $\rho_{\lambda,\theta,m}$, and $\tau_{\lambda,\theta,m}$ are the spectral directional reflectivity and transmissivity of the membrane, respectively. The characteristic matrix formalism (see Supplementary Note 2) [46] is used to calculate $\rho_{\lambda,\theta,m}$ and $\tau_{\lambda,\theta,m}$, and it can be applied to both thin films and multilayers. In the presence of a mirror backside (i.e., best-case scenario), the membrane is treated as a multilayer of 90 nm SiN, 1.4 mm vacuum (i.e., the space between the membrane and the optical fiber holder), and an aluminum mirror substrate. This case is referred to as the multilayer model (see Fig. 2b). In the presence of a blackbody backside surrounding (i.e., worst-case scenario), the membrane is treated as a standalone 90 nm SiN thin film. This case is referred to as the thin film model (see Fig. 2b). For conciseness, the spectral hemispherical emissivity ($\varepsilon_{\lambda,m}$) of the thin film model is shown in Fig. 1a, and the spectral hemispherical emissivity of the multilayer model is shown in Supplementary Note 3 (see supplementary Note 4 for $\varepsilon_{\lambda,m}$ calculation details).

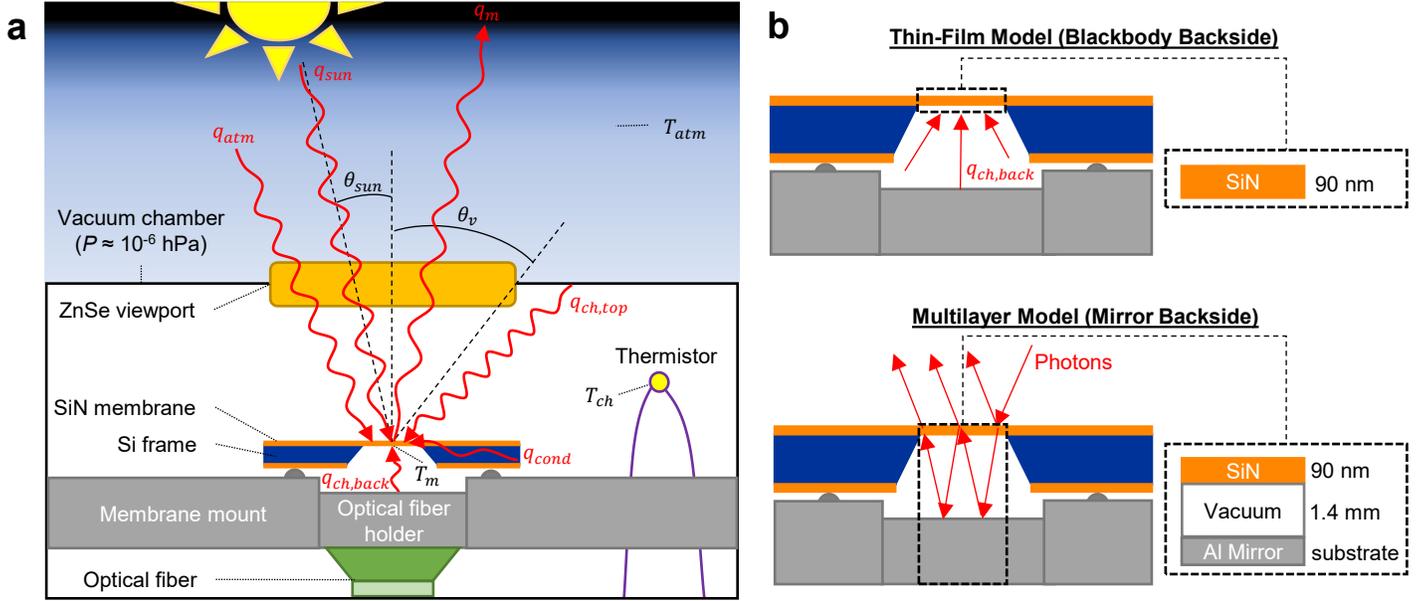

*Fig. 2. a) Radiative cooling experiment depiction: a freestanding SiN membrane mounted inside a vacuum chamber and facing the sky through a ZnSe viewport. b) Membrane emissivity models.*

### 2.2 Heat Transfer Rates

The rate at which radiation is leaving the membrane (in W/m²) is calculated using the total emissive power relation [45], and is given by:

$$q_m(T_m) = A_{corr} \int_0^\infty \int_0^{2\pi} \int_0^{\frac{\pi}{2}} \varepsilon_{\lambda,\theta,m}(\lambda,\theta) I_{BB}(\lambda,T_m) \cos\theta \, d\omega \, d\lambda, \tag{3}$$

where $d\omega = \sin\theta \, d\theta \, d\phi$ is the differential solid angle, $\lambda$, $\theta$, and $\phi$ are the wavelength, azimuth angle, and zenith angle of the radiation, respectively, and $\varepsilon_{\lambda,\theta,m}$ is the spectral directional emissivity of the membrane. The spectral radiative blackbody intensity ($I_{BB}$) is given by:

$$I_{BB}(\lambda,T) = \frac{2hc_0^2}{\lambda^5 \left( e^{\frac{hc_0}{\lambda k_B T}} - 1 \right)}, \tag{4}$$

where $h = 6.626 \times 10^{-34}$ J·s is the Planck constant, $k_B = 1.381 \times 10^{-23}$ J/K is the Boltzmann constant, $c_0 = 2.998 \times 10^8$ m/s is the speed of light in vacuum, and $T$ is the temperature of the blackbody [45]. The correction factor $A_{corr}$ accounts for the fact the thin film model ($A_{corr} = 2$) allows membrane radiation from both sides of the membrane, while the multilayer model ($A_{corr} = 1$) has a mirror backside (see Fig. 2b) that restricts radiation to the top side only.

Atmospheric irradiation is incident on the membrane only in the viewport range (i.e., $0 < \theta < \theta_v$), and the rate at which it is incident on the membrane is given by:

$$q_{atm}(T_{atm}) = \int_0^\infty \int_0^{2\pi} \int_0^{\theta_v} \tau_{\lambda,v}(\lambda) \varepsilon_{\lambda,\theta,atm}(\lambda,\theta) \alpha_{\lambda,\theta,m}(\lambda,\theta) I_{BB}(\lambda,T_{atm}) \cos\theta \, d\omega \, d\lambda, \tag{5}$$

where $\tau_{\lambda,v}$ is the spectral transmissivity of the anti-reflection coated ZnSe viewport at normal incidence (Thorlabs WG71050-E3 [44]; see Fig. 1a), and $\varepsilon_{\lambda,\theta,atm}$ is the spectral directional emissivity of the atmosphere. The ZnSe viewport is chosen for its selective transparency in the 8 μm – 13 μm wavelength range, thus minimizing parasitic heating from the atmosphere and the sun while not affecting the radiative cooling of the membrane. The spectral directional emissivity of the atmosphere has been modelled extensively in prior work, and Zhao et. al provided a detailed summary in [30]. A widely accepted [7], [23], [47]–[50] correlation for atmospheric emissivity is given by:

$$\varepsilon_{\lambda,\theta,atm}(\lambda,\theta) = 1 - \tau_{\lambda,atm}(\lambda)^{1/\cos\theta}, \tag{6}$$

where $\tau_{\lambda,atm}$ is the atmospheric spectral transmissivity at normal incidence.

The atmospheric transparency—which is maximum in the 8 μm – 13 μm range—is the critical component that makes radiative cooling possible by allowing a significant portion of radiation to escape the Earth towards outer space. There are many factors that affect $\tau_{\lambda,atm}$; the most important ones are the atmospheric water content [51], cloudiness [52], and location on Earth [53]. The Moderate Resolution Atmospheric Transmission (MODTRAN) model analyzes optical measurements through the atmosphere and is widely used to extract $\tau_{\lambda,atm}$ for various climate conditions [42], [43]. In this work, the experiments are conducted in the region of Ottawa, Canada, which is subject to variable atmospheric conditions. To account for this variability in the heat transfer model, $\tau_{\lambda,atm}$ is approximated for mid-latitude summer climate conditions as a weaker estimate, and for sub-arctic winter climate conditions as a stronger estimate (see Fig. 3a).

In the thin film model, the backside surrounding (i.e., the optical fiber holder and membrane mount) is treated as a large isothermal surrounding. As such, the chamber irradiation incident on the backside of the membrane can be approximated by emission from a blackbody ($\varepsilon = 1$) at the chamber temperature ($T_{ch}$) as:

$$q_{ch,back}(T_{ch}) = \int_0^\infty \int_0^{2\pi} \int_0^{\frac{\pi}{2}} \alpha_{\lambda,\theta,m}(\lambda,\theta) I_{BB}(\lambda,T_{ch}) \cos\theta \, d\omega d\lambda \, . \tag{7}$$

The multilayer model considers $q_{ch,back} = 0$ since the backside surrounding of the membrane is treated as an aluminum mirror that radiates no heat to the membrane.

The chamber irradiation incident on the topside of the membrane is given by:

$$q_{ch,top}(T_{ch}) = \int_0^\infty \int_0^{2\pi} \int_0^{\frac{\pi}{2}} \varepsilon_{\lambda,\theta,ch}(\lambda,\theta) \alpha_{\lambda,\theta,m}(\lambda,\theta) I_{BB}(\lambda,T_{ch}) \cos\theta \, d\omega d\lambda, \tag{8}$$

where the spectral directional emissivity of the surrounding topside chamber ($\varepsilon_{\lambda,\theta,ch}$) is angle dependent since the topside of the membrane can be coupled either to the chamber walls or the ZnSe viewport. Outside of the viewport range (i.e., $\theta_v < \theta < \pi/2$), the opaque chamber walls can be treated as a blackbody ($\varepsilon_{\lambda,\theta,ch} = 1$) because photons leaving the membrane have a negligible chance of leaving the chamber, i.e., they will reflect on the chamber walls until absorbed. Within the viewport range (i.e., $0 < \theta < \theta_v$), the radiation leaving the membrane and reflected by the viewport is assumed to have no chance of eventually exiting the chamber for the same reason. In other words, photons emitted by the membrane within the viewport range (i.e., $0 < \theta < \theta_v$) are either absorbed by the chamber or transmitted through the viewport directly after incidence from the membrane. As such, the angle dependent spectral emissivity of the chamber is given by:

$$\varepsilon_{\lambda,\theta,ch}(\lambda,\theta) = \begin{cases} 1 - \tau_{\lambda,v}(\lambda), & 0 < \theta < \theta_v \\ 1, & \theta_v < \theta < \frac{\pi}{2}. \end{cases} \tag{9}$$

Solar radiation reaching the membrane can be categorized into two components: (i) direct solar radiation, and (ii) diffuse solar radiation (i.e., atmospherically scattered radiation) [45]. The amount of each solar radiation component reaching the membrane is greatly affected by the atmospheric conditions, the geographical location, and the time of day and year. The AM1.5 spectrum (see Fig. 3b) is a commonly accepted international standard for the global (i.e., sum of direct and diffuse) solar intensity [54]. It is widely used for characterizing the performance of solar panels, and for characterizing the global spectral solar irradiance in the context of radiative cooling [23], [30], [49], [53]. The net solar irradiance of the AM1.5 spectrum is 1000 W/m², which is representative of the peak conditions in the Ottawa, Canada, region (see Supplementary Note 5). As such, global solar irradiation incident on the membrane can be calculated as:

$$q_{sun} = \int_0^\infty \tau_{\lambda,v}(\lambda) \alpha_{\lambda,\theta,m}(\lambda,\theta_{sun}) I_{AM1.5}(\lambda) d\lambda, \tag{10}$$

where $\theta_{sun}$ is the fixed angle at which the sun is facing the membrane, and depends on latitude, time of day, and day number of the year (see Supplementary Note 6 for $\theta_{sun}$ calculation details).

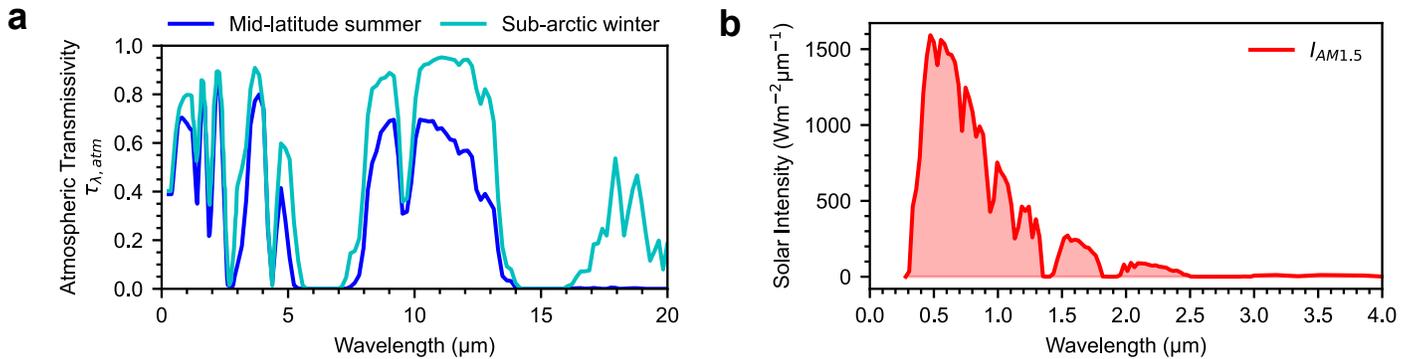

*Fig. 3. a) Atmospheric transmissivity at normal incidence for mid-latitude summer and sub-arctic winter climate conditions computed by the MODTRAN model [42], [43]. b) AM1.5 solar intensity spectrum, representing the distribution of solar power (in watts per square meter per nanometer of bandwidth) as a function of wavelength [54].*

Conductive heat transfer from the silicon frame to the SiN membrane depends on the membrane geometry and was modeled recently in two different studies [36], [38]. Considering that only conductive and radiative heat transfer occurs on the membrane in this work (i.e., no convection in high vacuum), the rate of conduction from the silicon frame to the SiN membrane can be calculated as:

$$q_{cond} = x_{cond} \cdot \frac{2\varepsilon_m \sigma (T_{ch}^4 - T_m^4)}{1 - x_{cond}}, \tag{11}$$

where $\varepsilon_m$ is the total hemispherical emissivity of the membrane (e.g., $\varepsilon_m = 0.097$ at $T_m = 300$ K in the thin film model), and $\sigma = 5.67 \times 10^{-8}$ W/m$^2$K is the Stefan-Boltzmann constant. Since $T_m$ is unknown, and since $\varepsilon_m$ was shown to vary negligibly between 200 K to 300 K in [36], we assume $T_m \approx T_{ch}$ when evaluating $\varepsilon_m$ (see Supplementary Note 4 for $\varepsilon_m$ calculation details). In turn, $x_{cond}$ is dependent on a membrane geometry factor ($\beta r_{eff}$) given by [36]:

$$x_{cond} = \frac{2}{\beta r_{eff}} \frac{I_1(\beta r_{eff})}{I_0(\beta r_{eff})}, \tag{12}$$

where $I_N$ is the $N$th-order modified Bessel function of the first kind. $\beta r_{eff}$ is defined as [36]:

$$\beta r_{eff} = \left( \sqrt{\frac{8\sigma \varepsilon_m T_{ch}^3}{kd}} \right) \cdot \left( 1.252 \frac{L}{2} \right), \tag{13}$$

where $k$ is the thermal conductivity of the membrane ($k \approx 2.7$ W/m·K for SiN [37]), $d \approx 90$ nm is the thickness of the membrane, and $L \approx 6$ mm is the side length of the square membrane.

## 3    Experimental Method

We assess the temperature drop of a freestanding SiN membrane upon exposure to the sky by tracking its mechanical resonance frequency with the experimental setup depicted in Fig. 4a (schematic) and Fig. 4b (picture). The experimental setup includes four main systems: an optical fiber interferometer for vibration measurements, the vacuum chamber, the power system, and the data acquisition (DAQ) system. Each system is described below.

A conventional optical fiber interferometer (see Fig. 4a) [55] is used to measure the resonance frequency shift of the membrane upon radiative cooling. The temperature drop of the SiN membrane during radiative cooling results in an increase in membrane tensile stress, which leads to a proportional rise in resonance frequency [56]–[58]. Zhang et al. [36] demonstrated that, for a membrane at uniform temperature, the membrane average temperature shift can be approximated from the resonance frequency shift ($\Delta f_m$) by:

$$\overline{\Delta T_m} = -[1 + e] \frac{2\varsigma_0 (1 - \nu)}{a_{CTE} E f_0} \Delta f_m, \tag{14}$$

where $a_{CTE}$ is thermal expansion coefficient, $E$ is Young's modulus, $f_0$ is the initial resonance frequency (i.e., before the frequency shift), and $\nu$ is the Poisson ratio of the SiN membrane. The initial (i.e., before the frequency shift) membrane tensile stress ($\varsigma_0$) can be calculated from the mass density ($\rho$), the side length ($L$), and the number of antinodes in the vibrational mode shape ($m, n$) of the membrane using the following relation [59]:

$$\varsigma_0 = \frac{4f_0^2 \rho L^2}{(m^2 + n^2)}. \tag{15}$$

According to Zhang et. al [36], the approximation that the temperature profile of the membrane is uniform leads to an underestimation error of the membrane temperature shift, which can be compensated with a correction factor ($e \leq 0.2$) obtained from finite element simulations. The value of $e$ is dependent on $\beta r_{eff}$ and $(m, n)$ and can be obtained from Fig. 4 in [36]. In this work, the resonance frequency of the vibrational mode shape ($m = 1$, $n = 1$) is tracked, which results in a correction factor $e \approx 0.12$ (for $T_{ch} = 300$ K).

The membrane dimensions were measured in [37] yielding thicknesses ($d$) of 90 ± 1.7 nm and side lengths ($L$) of 6 ± 0.015 mm. A description of the membrane fabrication process from low stress, low pressure chemical vapor deposition (LPCVD), SiN is available in [60]. The material properties of low stress LPCVD SiN are obtained from literature [61]–[66] as $a_{CTE} = 2.2 \pm 0.1 \times 10^{-6}$ K$^{-1}$, $E = 290 \pm 90$ GPa, $\nu = 0.25 \pm 0.03$, and $\rho = 3000 \pm 100$ kg/m$^3$. The uncertainties of the material properties lead to an uncertainty of $\overline{\Delta T_m}$ in eq. (14), which is calculated using the propagation of uncertainty rule (see Supplementary Note 7).

Vacuum environments are often used in optomechanical studies to remove convective heat transfer effects and, in the case of mechanical resonators, prevent overdamping of the membrane vibrations. For this work, the SiN membrane is placed inside a vacuum chamber kept at high vacuum ($P \approx 10^{-6}$ hPa) with an ion pump. The membrane is mounted inside the nipple of the vacuum chamber as closely as possible to the anti-reflection coated ZnSe viewport (Thorlabs WG71050-E3 [44]) to maximize the field of view of the membrane (see Fig. 4c). The stainless-steel membrane mount is anchored to the chamber walls with spring plungers (see Fig. 4d). Two different methods were used to fix the membrane on the mount: (i) magnetic spheres, (ii) vacuum-compatible tape. In the first method, the membrane is placed between three sets of magnetic spheres to hold it in a fixed position on the mount (see Fig. 4d). This method is optimal for higher quality factor measurements—i.e., less dissipation, meaning longer vibrations—however, it is not strong enough to hold the membrane in place when a resistance temperature detector (RTD) is attached to the Si frame (as shown in Fig. 4c). As such, the second method is applied for measurements with the RTD, where vacuum compatible tape is used to anchor the membrane on the mount (see Fig. 4c). In all cases, the position of the membrane is chosen to ensure an optimal alignment with the optical fiber tip and the viewport. The optical fiber inside the chamber is connected to the optical fiber interferometer (outside the chamber) via the optical feedthrough. A vacuum chamber holder (see Fig. 4b) is designed to hold the chamber in a position which ensures that the membrane is facing the sky, and adjustable glides allow the levelling of the holder to ensure the membrane is pointing to the zenith.

The power system is comprised of two 12 V batteries and a power distribution enclosure (see Fig. 4a) to ensure the portability of the system. It provides power to the ion pump controller (24 V input voltage) and the optical fiber interferometer (24 V input voltage). See Supplementary Note 8 for details on the power distribution enclosure. A portable computer provides power to the oscilloscope and DAQ device, and it is also used for data storage and processing.

The DAQ system uses a DAQ device (LabJack U6-Pro) to continuously measures the ambient atmospheric temperature ($T_{atm}$) with a temperature probe (LabJack EI-1034) located outside of the chamber (see Fig. 4b), the temperature inside the vacuum chamber ($T_{ch}$) with a thermistor (Measurement Specialties 55036 Precision Glass NTC) located inside the nipple of the chamber (see Fig. 4c), and the silicon frame temperature ($T_{Si}$) using an RTD (TE Connectivity 1 kOhm PTF) silver pasted to the Si frame (see Fig. 4c).

The following procedure is applied for each radiative cooling experiment: (i) bring the experimental setup (shown in Fig. 4b) outside with a clear view of the sky and begin continuously recording $T_{ch}$, $T_{atm}$, and $T_{Si}$, (ii) wait for $T_{ch}$, $T_{atm}$, and $T_{Si}$ to stabilize (typically 15 minutes, depending on outdoor temperature and sun exposure), (iii) cover the ZnSe viewport with aluminum foil to ensure the membrane is coupled to the chamber only, i.e., no radiative cooling occurs, (iii) excite the membrane by light mechanical impact on the outer wall of the vacuum chamber nipple—i.e., to make the membrane vibrate—and begin continuously tracking the resonance frequency of the membrane ($f_m$), (iv) uncover the ZnSe viewport to expose the membrane to the sky—i.e., for radiative cooling to occur—and measure the membrane resonance frequency shift ($\Delta f_m$), (v) calculate the membrane temperature drop from $\Delta f_m$ using eq. (14).

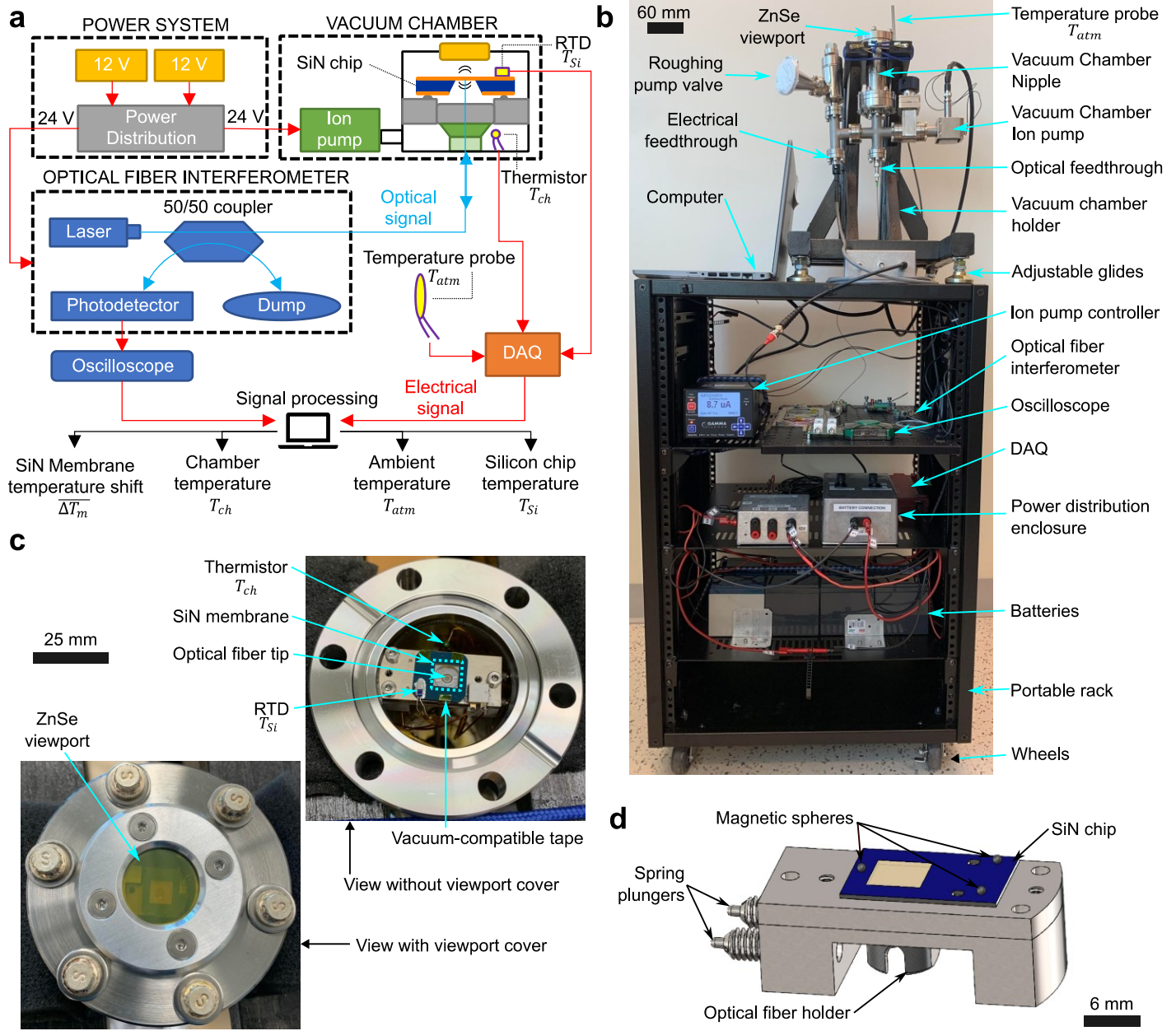

Fig. 4. Experimental setup a) schematic and b) picture. c) Top view of the vacuum chamber with, and without, the viewport cover. d) Membrane mount design.

## 4    Results & Discussion

We assess the temperature drop of a freestanding SiN membrane in vacuum upon radiative cooling for various experimental conditions (i.e., date, time, location, cloudiness, relative humidity, $T_{ch}$, $T_{atm}$, $\theta_v$, and $\theta_{sun}$). Experiments were conducted at three different locations: (i) at the front entrance of the Advanced Research Complex at the University of Ottawa (uOttawa), (ii) in an open field in the rural region of Vankleek Hill, ON, Canada, and (iii) in an open field in the rural region of Fournier, ON, Canada. Fig. 5 displays pictures of the experimental setup in each location. Experiments (20 in total) were conducted on 15 different days during fall and summer months, resulting in successful radiative cooling of the membrane in each case. The experiment conditions (i.e., date, time, location, cloudiness, relative humidity, $T_{ch}$, $T_{atm}$, $\theta_v$, and $\theta_{sun}$) and results, along with pictures of the sky during each experiment, are provided in Supplementary Note 9.

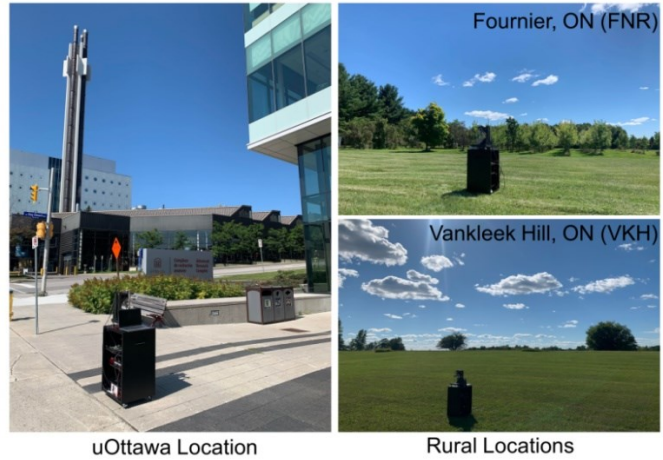

*Fig. 5. Pictures of the experiment locations. Left image: front of the Advanced Research Complex at the University of Ottawa, ON, Canada. Right top image: open field in the rural region of Vankleek Hill, ON, Canada. Right bottom image: open field in the rural region of Fournier, ON, Canada.*

Fig. 6a presents an experiment confirming that a temperature gradient on-chip between the SiN membrane and the silicon frame is created via radiative cooling. The reduction in temperature of the SiN membrane stabilizes approximately one second after uncovering the viewport, while the silicon frame remains approximately at constant temperature. The silicon frame is eventually heated by exposure to sunlight, but this occurs on a much larger time scale (see Supplementary Note 10). It is possible that heating of the silicon frame can enhance the magnitude of the thermal gradient since thermal coupling of the membrane is radiation-dominated (i.e., it is weakly affected by heat conduction from the frame). However, this idea is not explored further, and we limit experiments to short time scales during which the silicon frame temperature remains constant.

Successful cooling is achieved in all experiments (20 in total), and we observe that atmospheric conditions (i.e., clouds, humidity) are the main factors responsible for the variability of the cooling, while the SiN membrane is almost entirely unaffected by solar irradiation. Fig. 6b compiles all experimental results compared with the corresponding theoretical results (i.e., for the respective $T_{ch}$, $T_{atm}$, $\theta_v$, and $\theta_{sun}$). The heat transfer model bounds account for the two membrane emissivity models, and for the variability of the atmospheric transparency. The weak cooling bound is calculated using the weaker thin film model emissivity (see Supplementary Note 3) and the weaker mid-latitude summer atmospheric transparency (see Fig. 3a), and the strong cooling bound is calculated using the stronger multilayer model emissivity (see Supplementary Note 3) and the stronger sub-arctic winter atmospheric transparency (see Fig. 3a). However, the model bounds account for clear sky and clear field of view conditions, which explains why some cloudy experiments show weaker cooling than expected by the model (e.g., 05-20-2022 and 08-15-2022). It is evident that cloudiness and sky transparency conditions play a significant role in the magnitude of the membrane temperature drop. For example, two experiments were conducted 20 minutes apart on 05-11-2022 (at 22:10 and 22:30) while a cloud was passing over the setup (see pictures of the sky in Supplementary Note 9). This resulted in a 1.8 K (roughly 50%) increase in temperature drop from cloudy ($3.7 \pm 1.2$ K drop) to clear sky ($5.5 \pm 1.7$ K drop) conditions. The same effect occurred on 08-27-2022 between 13:00 and 13:20, where a 2.2 K (roughly 50%) increase in temperature drop was observed from cloudy ($4.3 \pm 1.4$ K drop) to clear sky ($6.5 \pm 2.1$ K drop) conditions. Moreover, it is evident that the urban landscape at the uOttawa location affects the cooling of the membrane because the rural experiments—with open field of view conditions—obtained significantly better results. This is most likely due to field of view obstruction as well as parasitic heating from the buildings and streetlights at the uOttawa location.

The maximal temperature drops were observed at the rural locations, with clear sky conditions, on 08-13-2022. Three different experiments were conducted on this day and resulted in the record daytime cooling ($9.3 \pm 2.9$ K at 16:30 with indirect sunlight, i.e., $\theta_{sun} > \theta_v$) and record nighttime cooling ($7.1 \pm 2.2$ K at 22:25). A daytime experiment with direct sunlight was also conducted ($9.0 \pm 2.8$ K at 13:45) and confirm the weak influence of sunlight on our membrane. We note that the relative humidity was significantly higher during the nighttime experiment than during the daytime experiments. This means the atmospheric transparency was most likely weaker, which would explain the weaker temperature drop during the nighttime experiment.

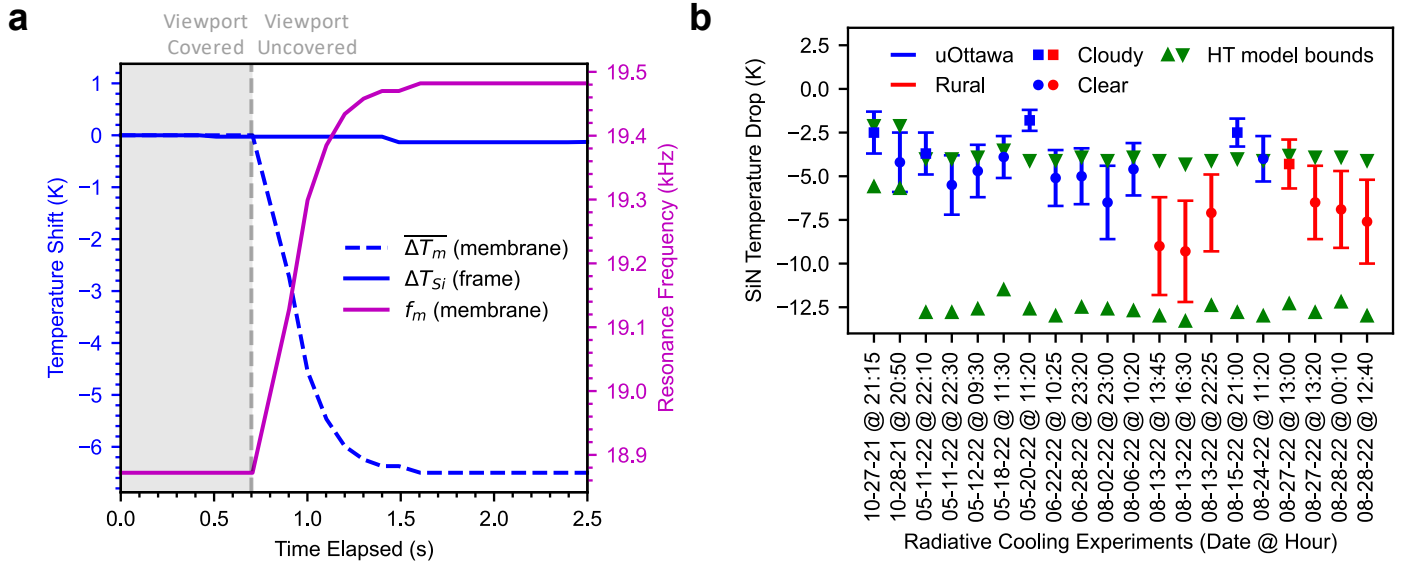

Fig. 6. a) Transient resonance frequency increase and temperature drop of the SiN membrane upon radiative cooling while the silicon frame remains at constant temperature for an experiment conducted on 08-27-2022 at 13:20, under direct sunlight. b) Membrane experimental temperature drop vs. theoretical temperature drop range for all experiments.

The parasitic heating sources (i.e., $q_{atm}$, $q_{ch,top}$, $q_{ch,back}$, $q_{cond}$, $q_{sun}$) can be minimized, or even eliminated, by improving the experimental setup. Fig. 7a shows the weight of each parasitic heating rate against $q_m$ in the steady state energy balance considering the thin film model emissivity, and Fig. 7b shows the theoretical cooling potential of the membrane after successive improvements to the experimental setup. The thin film model is used as the baseline. The first improvement is to maximize the field of view of the membrane (i.e., to $\theta_v = \pi/2$). This would eliminate the chamber walls portion of $q_{ch,top}$, and it could be achieved by using a larger viewport. The second improvement is to maximize backside reflection by approaching a perfect mirror with the backside surrounding of the membrane. This would eliminate $q_{ch,back}$, and it could potentially be achieved through polishing of the optical fiber holder surface, or adding a reflective layer (i.e., aluminum) under the SiN membrane. The third improvement is to eliminate conductive heating effects from the Si frame ($q_{cond}$). This could potentially be achieved by increasing the side length of the membrane ($L$), or by patterning the membrane in a trampoline shape [67], which would increase the fraction of heat transfer occurring via radiation versus conduction. The fourth, more hypothetical, improvement is to replace the SiN membrane with an ideal emitter—possessing an emissivity of unity in the 8 μm to 13 μm range and null elsewhere—to (i) maximize $q_m$ in the atmospheric transparency window, and (ii) eliminate $q_{sun}$ and a significant portion of $q_{atm}$. While achieving a perfect emitter is unlikely, emissivity engineering (e.g., with metasurfaces and multilayers) could potentially increase the emissivity peak of the SiN membrane in the 8 μm to 13 μm range. The final improvement is to replace the ZnSe viewport with an ideal viewport, which would be transparent at all wavelengths to fully eliminate $q_{ch,top}$. These modifications would leave $q_{atm}$ (in the 8 μm – 13 μm range only) as the only remaining parasitic heating source. Fig. 7b displays these improvements considering both mid-latitude summer and sub-arctic winter atmospheric transparency conditions. The maximal membrane temperature drop achieved with the current experimental setup is 9.3 K and, depending on atmospheric transparency conditions, it could potentially increase to 22 K – 48 K with the same membrane and the suggested setup improvements (i.e., improvements 1 to 3), and up to 30 K – 67 K with a theoretical ideal emitter and viewport. In the context of a heat engine on-chip, 48 K cooling corresponds to a maximum Carnot efficiency of approximately 16%. While this is lower than typical solar cells, it would have the advantage of working 24 hours a day.

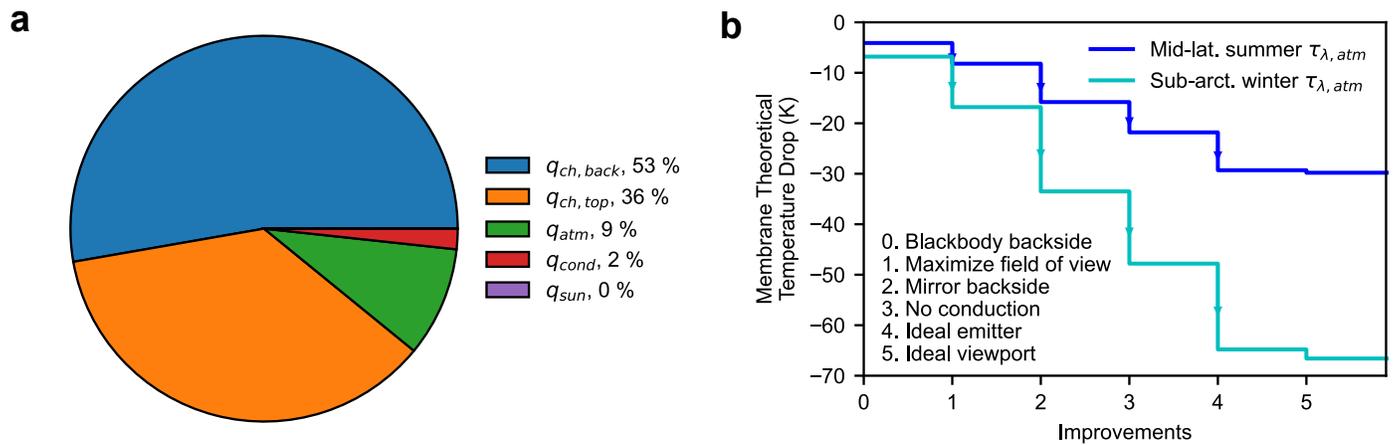

Fig. 7. a) Theoretical weight of each parasitic heating rate against $q_m$ in the steady state energy balance for the thin film model. b) Theoretical membrane temperature drops for successive improvements to the experimental setup (with constants $T_{ch} = T_{atm} = 300$ K, and $\theta_{sun} = 0$).

## 5    Conclusion

We demonstrated a localized thermal gradient on-chip by radiative cooling of a silicon nitride nanomechanical resonator while its supporting silicon frame remained at constant temperature. Successful cooling of the SiN membrane was demonstrated in various atmospheric conditions, and maximal temperature drops of $9.3 \pm 2.9$ K and $7.1 \pm 2.2$ K were demonstrated during the day and night, respectively. Improvements to the experimental setup could increase the temperature drop to, depending on atmospheric conditions, 22 K – 48 K for the same membrane, while emissivity engineering eventually yields a theoretical maximum cooling of 30 K – 67 K with an ideal emitter. This work is a first step towards creating sustainable micro heat engines on-chip for the purpose of self-powering in small scale applications.

### Acknowledgments

This work is supported by the Natural Sciences and Engineering Research Council of Canada (NSERC) Discovery Program [grant number: RGPIN-2018-04412] and the National Research Council of Canada (NRC) Ideation Fund: New Beginnings Initiative [grant number: INB-000574-1].

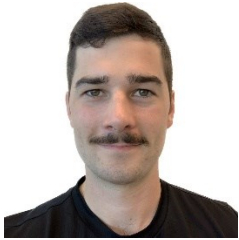

**Mr. Alexandre Bouchard** received his B.A.Sc. degree in Mechanical Engineering from the Department of Mechanical Engineering at the University of Ottawa in 2020. He is currently pursuing a M.A.Sc. degree in Mechanical Engineering at the University of Ottawa under the supervision of Prof. Raphael St-Gelais, for which he received the NSERC CGS-M scholarship in 2022. His current research focuses on the cooling of nanostructures below ambient temperature by radiative coupling to outer space.

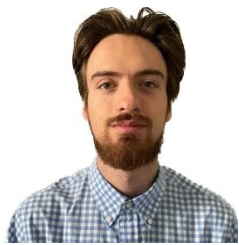

**Mr. Timothy Hodges** received his B.A.Sc. degree in Mechanical Engineering from the Department of Mechanical Engineering at the University of Ottawa in 2020. He is currently completing his M.A.Sc. degree in Mechanical Engineering at the University of Ottawa under the supervision of Prof. Raphael St-Gelais. His current research focuses on high precision inertial sensing using ultra high-quality factor nanomechanical resonators.

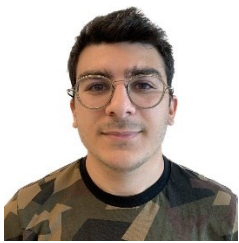

**Mr. Michel Stephan** received his B.A.Sc. degree in Mechanical Engineering from the Department of Mechanical Engineering at the University of Ottawa in 2021. He is currently pursuing a M.A.Sc. degree in Mechanical Engineering at the University of Ottawa under the supervision of Prof. Raphael St-Gelais. His current research focuses on the long-term stability performance of silicon nitride nanomechanical resonators.

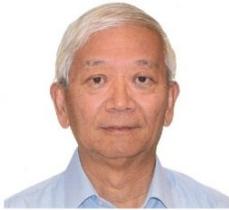

**Dr. Lixue Wu** is a research officer of the National Research Council Canada (NRC). He joined the NRC in 1994 after holding Natural Sciences and Engineering Research Council of Canada (NSERC) Postdoctoral Fellowship for two years and undertaking research that involves collaboration with other Canadian organizations. He obtained his Ph.D. from the University of Victoria in the Department of Electric and Computer Engineering in 1992. His research interests are mainly focused on measurement science and standards in the areas of acoustics, ultrasound and vibration. He is the chair of the Canadian mirror committees of the ISO/TC 108 and IEC/TC 29.

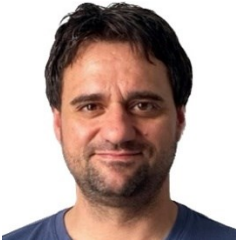

**Dr. Triantafillos Koukoulas** is a senior research officer at the Metrology Research Centre, National Research Council Canada. He obtained his B.Eng. from the University of Bradford in 1997, his M.Sc. from the University of Brighton and University of Sussex in 1998 and the Ph.D. from the University of Sussex in 2003. He is on the editorial board of the International Journal of Acoustics and Vibration, is a fellow of the Institute of Physics, senior member of the IEEE and member of the International Institute of Acoustics and Vibration. His research interests primarily focus on optical measurement techniques for acoustical metrology

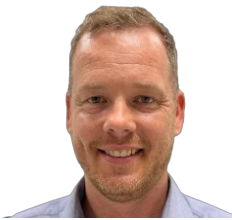

**Dr. Richard Green** is a senior research officer at the Metrology Research Centre of the National Research Council Canada. He obtained a B.Sc. in Physics, Mathematics and Statistics from Concordia University in Montreal, and PhD. from the university of Ottawa in Physical Chemistry. Since 2011 he has specialized in high precision mass metrology in air and vacuum, with research focus on high accuracy measurement of Planck constant and realization of the kilogram. He is vice chair of the Consultative Committee for Mass and Related Quantities working group on Mass and co-winner of the 2023 IEEE Joseph F. Keithley award.

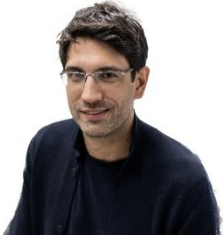

**Dr. Raphael St-Gelais** is an Assistant Professor in the Department of Mechanical Engineering at the University of Ottawa. He received his Ph.D. from Polytechnique Montreal (Engineering Physics; 2009 to 2012) where he worked on integrated optical sensors and MEMS-tunable optical components, for which he was awarded the Alexander Graham Bell Canada Graduate Scholarship. He received postdoctoral fellowships from FRQNT and NSERC for his work at Cornell & Columbia. He currently leads the uOMEMS lab at the University of Ottawa in research that focuses on the interplay of optical, thermal, and mechanical phenomena in custom made micro and nano scale devices.

**Supplementary Information**

# Supplementary Information: Localized Thermal Gradients On-Chip by Radiative Cooling of Silicon Nitride Nanomechanical Resonators

Alexandre Bouchard[a], Timothy Hodges[b,c], Michel Stephan[a], Lixue Wu[b], Triantafillos Koukoulas[b], Richard Green[b], Raphael St-Gelais[a,c,d]

[a] Department of Mechanical Engineering, University of Ottawa, 75 Laurier Avenue East, Ottawa, Ontario K1N 6N5, Canada
[b] Metrology Research Centre, National Research Council Canada, 1200 Montreal Road, Ottawa, Ontario K1A 0R6, Canada
[c] Centre for Research in Photonics, University of Ottawa, 75 Laurier Avenue East, Ottawa, Ontario K1N 6N5, Canada
[d] Department of Physics, University of Ottawa, 75 Laurier Avenue East, Ottawa, Ontario K1N 6N5, Canada

## SUPPLEMENTARY INFORMATION TABLE OF CONTENTS

# S1.   CONVECTIVE EFFECTS AT HIGH-VACUUM

The convection heat transfer between the membrane and the ambient air in the chamber is dependent on the gas dynamics in the chamber, which can be characterized by the Knudsen number:

$$K_n = \frac{\Lambda}{l},$$

(S1)

where $\Lambda$ is the mean free path of the air molecules and $l$ is the characteristic length (i.e., the gaseous gap distance over which thermal transport occurs) [1]. Here, there are two characteristic lengths: (i) the backside gap (i.e., the distance between the membrane and the optical fiber holder, $l \sim 1.4$ mm), and (ii) the topside gap (i.e., the distance between the membrane and the viewport, $l \sim 1.0$ cm). From the ideal gas law, the mean free path of air can be determined as:

$$\Lambda = \frac{k_B T}{\sqrt{2}\pi P D_m^2},$$

(S2)

where $P$ and $T$ are the gas pressure and temperature, respectively, $k_B = 1.381 \times 10^{-23}$ J/K is the Boltzmann constant, and $D_m \approx 0.3$ nm is the molecular diameter of air. For a chamber at $P \approx 1 \times 10^{-6}$ hPa and $T \approx 300$ K, the mean free path is approximately $\Lambda \approx 104$ m, which corresponds to a Knudsen number of $K_n \approx 74{,}000$ for the backside and $K_n \approx 10{,}400$ for the topside. Since the mean free path of the air molecules is much larger than the characteristic lengths in the chamber (i.e., $K_n \gg 10$), the air flow in the chamber may be defined by the "free molecule flow" regime [1]. The air conduction net heat flux between two plates (i.e., the membrane and the chamber in this case) in the free molecule regime can be determined by [1]:

$$q_{air,FM} = \frac{T_m - T_c}{\frac{(2 - A_T)\sqrt{8\pi R_{air} T_{m,FM}}}{A_T(\gamma + 1)c_v P}},$$

(S3)

where $T_m$ and $T_c$ are the temperatures of the membrane and chamber, respectively, $A_T$ is the thermal accommodation coefficient, $R_{air} = 287$ J·kg$^{-1}$K$^{-1}$ is the gas constant of air, $\gamma = 1.4$ is the specific heat ratio for diatomic gases, $c_v = 716.6$ J·kg$^{-1}$K$^{-1}$ is the constant volume specific heat for air, and $T_{m,FM}$ is the effective mean temperature for a free molecule flow regime [1]. The thermal accommodation coefficient ($A_T$) is dependent on various parameters (e.g., materials, surface conditions, temperature, etc.) and has been measured experimentally for a wide range of conditions [2]. For a wide range of engineered surfaces, the thermal accommodation coefficient is $0.8 < A_T < 1$. For this analysis, the coefficient will be taken as $A_T = 0.9$ (i.e., the coefficient for flow of air with glass in [2]). The effective mean temperature for a free molecule regime is given by:

$$T_{m,FM} = \frac{4T_m'T_c'}{\left(\sqrt{T_m'} + \sqrt{T_c'}\right)^2}, \tag{S4}$$

where,

$$T_m' = \frac{T_m + (1 - A_T)T_c}{2 - A_T}, \tag{S5}$$

$$T_c' = \frac{T_c + (1 - A_T)T_m}{2 - A_T}. \tag{S6}$$

By taking $P$ = 1 × 10⁻⁶ hPa, $T_c$ = 300 K, and $T_m$ = 290 K (i.e., assuming a 10 K membrane temperature drop), the air conduction net heat flux on the membrane inside the chamber is $q_{air,FM}$ = 1.61 × 10⁻⁴ W/m². Considering the membrane radiative power ($q_m$) is in the 40 W/m² range (calculated with eq. (3) of main text), the air conduction only offsets around 0.0004% of the cooling power. Thus, the convection, or air conduction, effect on the membrane in the chamber at high vacuum is negligible.

## S2.   CHARACTERISTIC MATRIX FORMALISM

In the characteristic matrix formalism [3], the spectral directional reflectivity ($\rho_{\lambda,\theta,m}$) and transmissivity ($\tau_{\lambda,\theta,m}$) of a thin film assembly are a function of the admittance of the incident medium ($\eta_{inc}$) and output medium or substrate ($\eta_{out}$), and the characteristic matrix ($B, C$), and are calculated as:

$$\rho_{\lambda,\theta,m} = \left(\frac{\eta_{inc}B - C}{\eta_{inc}B + C}\right)\left(\frac{\eta_{inc}B - C}{\eta_{inc}B + C}\right)^*, \tag{S7}$$

$$\tau_{\lambda,\theta,m} = \frac{4\eta_{inc}Re(\eta_{out})}{(\eta_{inc}B + C)(\eta_{inc}B + C)^*}. \tag{S8}$$

The characteristic matrix ($B, C$) of the thin film assembly of $z$ layers is a function of the refractive index ($N_r$), thickness ($d_r$), and propagation angle ($\theta_r$) of each layer, along with the wavelength ($\lambda$) and the output medium admittance ($\eta_{out}$), and is calculated as:

$$\begin{bmatrix} B \\ C \end{bmatrix} = \left\{\prod_{r=1}^{z}\begin{bmatrix} cos\delta_r & (isin\delta_r)/\eta_r \\ i\eta_r sin\delta_r & cos\delta_r \end{bmatrix}\right\}\begin{bmatrix} 1 \\ \eta_{out} \end{bmatrix}, \tag{S9}$$

where,

$$\delta_r = \frac{2\pi N_r d_r cos\theta_r}{\lambda}, \tag{S10}$$

$$\eta_{r,TE} = N_r cos\theta_r \text{ for s-polarisation (transverse-electric)}, \tag{S11}$$

$$\eta_{r,TM} = N_r/cos\theta_r \text{ for p-polarisation (transverse-magnetic)}. \tag{S12}$$

Since the wave vector in the plane of incidence is constant throughout the assembly of thin films, the cosine of the propagation angle may be substituted with the following equation (derived from Snell's law and trigonometric functions):

$$cos\theta_r = \sqrt{1 - \left(\frac{N_{inc}sin\theta_{inc}}{N_r}\right)^2}, \tag{S13}$$

where $\theta_{inc}$ and $N_{inc}$ are the incident angle and refractive index, respectively. Using the previous equations, the s-polarized and p-polarized $\rho_{\lambda,\theta,m}$ and $\tau_{\lambda,\theta,m}$ are be determined, and the average of the real parts of both polarizations is taken to get the unpolarized $\rho_{\lambda,\theta,m}$ and $\tau_{\lambda,\theta,m}$. Note that the spectral refractive index of SiN is inferred from the Maxwell-Helmholtz-Drude model developed in [4].

## S3.   MEMBRANE EMISSIVITY MODELS

The heat transfer model accounts for two different membrane emissivity models: (i) The thin-film model, which treats the membrane as a 90 nm thick SiN membrane, and (ii) the multilayer model, which treats the membrane as a multilayer of 90 nm SiN, 1.4 mm vacuum, and an aluminum mirror (see Fig.2b in the main text). The spectral hemispherical emissivity spectrums of both models are shown below.

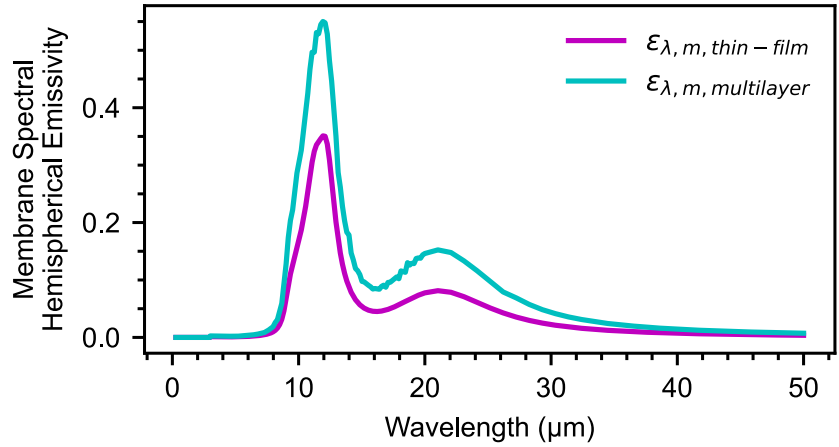

*Fig. S1. Membrane spectral hemispherical emissivity of the thin-film model vs. the multilayer model.*

## S4.    MEMBRANE EMISSIVITY CALCULATIONS

Knowing the directional spectral emissivity of the membrane ($\varepsilon_{\lambda,\theta,m}$; see eq. (2) of main text), the spectral hemispherical emissivity (i.e., integrated over all angles) of the membrane is calculated as [5]:

$$\varepsilon_{\lambda,m}(\lambda) = 2 \int_0^{\frac{\pi}{2}} \varepsilon_{\lambda,\theta,m}(\lambda, \theta)\, \cos\theta \, \sin\theta d\theta. \tag{S14}$$

In turn, the total hemispherical emissivity (i.e., integrated over all angles and wavelengths) of the membrane at temperature $T_m$ is calculated as [5]:

$$\varepsilon_m = \frac{\int_0^\infty \varepsilon_{\lambda,m}(\lambda) q_{\lambda,BB}(\lambda, T_m)}{\int_0^\infty q_{\lambda,BB}(\lambda, T_m)}, \tag{S15}$$

where,

$$q_{\lambda,BB}(\lambda, T_m) = \pi I_{BB}(\lambda, T_m). \tag{S16}$$

## S5.    OTTAWA SOLAR IRRADIANCE

The PVWatts calculator by the National Renewable Energy Laboratory [6] provides solar irradiation estimates for solar panels based on typical weather data from a multi-year historical period. Table S1 outlines the parameters used to determine the solar irradiance on the SiN membrane in this experiment.

*Table S1. PVWatts calculator parameters*

| Specification | Input |
|---|---|
| Requested Location | Ottawa, ON, Canada |
| Latitude | 45.41º North |
| Longitude | 75.7º West |
| Surface Tilt | 0º (horizontal) |

The calculator outputs an estimate of the hourly direct and diffuse solar irradiation (in $W/m^2$) components separately. By taking the sum of these two components, the global solar irradiance (in $W/m^2$) can be determined. For the parameters of Table S1, the monthly maximum global solar irradiation ($q_{sun,max}$) is outlined in Table S2. As shown, the peak solar irradiation conditions in Ottawa are in the 1000 $W/m^2$ range. This means the AM1.5 solar intensity spectrum is a good representation for the peak conditions in Ottawa.

*Table S2. Monthly maximum global solar irradiation on a horizontal surface in Ottawa*

| Month | 1 | 2 | 3 | 4 | 5 | 6 | 7 | 8 | 9 | 10 | 11 | 12 |
|---|---|---|---|---|---|---|---|---|---|---|---|---|
| $q_{sun,max}$ (W/m$^2$) | 1002 | 1096 | 1119 | 1104 | 1087 | 1056 | 1032 | 1026 | 1043 | 1014 | 979 | 949 |

## S6.    ANGLE AT WHICH THE SUN IS FACING THE MEMBRANE

Assuming the membrane can be treated as a horizontal surface on the surface of the Earth, the angle at which the sun is facing the membrane can be calculated as [7]:

$$\theta_{sun} = \cos^{-1}(\cos \Phi \cos \delta \cos \Omega + \sin \Phi \sin \delta),  \qquad \text{(S17)}$$

where $\Phi$ is the latitude, $\Omega$ is the hour angle, and $\delta$ is the declination angle. The hour angle ($\Omega$) is the angular displacement of the sun east or west of the local meridian due to the rotation of the earth, with 15º per hour (with afternoon being positive, and morning being negative). The declination angle is given by the equation of Cooper [7]:

$$\delta = 23.45 \sin \left( 360 \frac{284 + n}{365} \right)  \qquad \text{(S18)}$$

where $n$ is the day number of the year.

For example, for an experiment conducted on 2022-05-11, in Ottawa, at 9:30, $\theta_{sun}$ is determined with the following steps:

1. Ottawa has a latitude of 45º N, so $\Phi$ = 45º.

2. 2022-05-11 is the 131st day of the year, so $n$ = 131.

3. From eq. (S18) with $n$ = 131, $\delta$ = 17.8º.

4. 9:30 is 2.5 hours before noon, so $\Omega$ = -2.5 hours × 15º/hour = -37.5º.

5. From eq. (S17), $\theta_{sun}$ = 41.4º or 0.72 rad.

## S7.   TEMPERATURE SHIFT CALCULATION UNCERTAINTY

The membrane temperature shift ($\overline{\Delta T_m}$) calculation in eq. (14) of the main text is dependent on material properties with uncertainties. We use the propagation of uncertainty rule to determine the temperature shift uncertainty ($u_{\overline{\Delta T_m}}$) as:

$$
\begin{aligned}
&u_{\overline{\Delta T_m}} \\
&= \sqrt{\left(\frac{\partial \overline{\Delta T_m}}{\partial a_{CTE}} u_{a_{CTE}}\right)^2 + \left(\frac{\partial \overline{\Delta T_m}}{\partial L} u_L\right)^2 + \left(\frac{\partial \overline{\Delta T_m}}{\partial a_E} u_E\right)^2 + \left(\frac{\partial \overline{\Delta T_m}}{\partial \nu} u_\nu\right)^2 + \left(\frac{\partial \overline{\Delta T_m}}{\partial \rho} u_\rho\right)^2},
\end{aligned}
\qquad \text{(S19)}
$$

where $\frac{\partial \overline{\Delta T_m}}{\partial x}$ represents the partial derivative of $\overline{\Delta T_m}$ with respect to the material property $x$, and $u_x$ is the uncertainty of the material property $x$. See section 3 of the main text for the values and uncertainties of each material property.

# S8.   POWER DISTRIBUTION ENCLOSURE

A power distribution enclosure was designed to distribute the required voltage from two 12 V batteries to the optical fiber interferometer (24 V) and ion pump controller (24 V), and to ensure the safety of the equipment and users (see Fig. S2). There are three electronic components included in the circuit: a diode, switches, and fuses. A diode only allows current to flow from the anode to the cathode, so it ensures the correct polarity of the battery connections to the enclosure. Switches are used to allow current flow to the electronic components only when needed. Fuses ensure that the appropriate amount of current reaches the electronic components. If too much current is supplied by the batteries, the fuses will break the circuit to ensure the safety of the equipment.

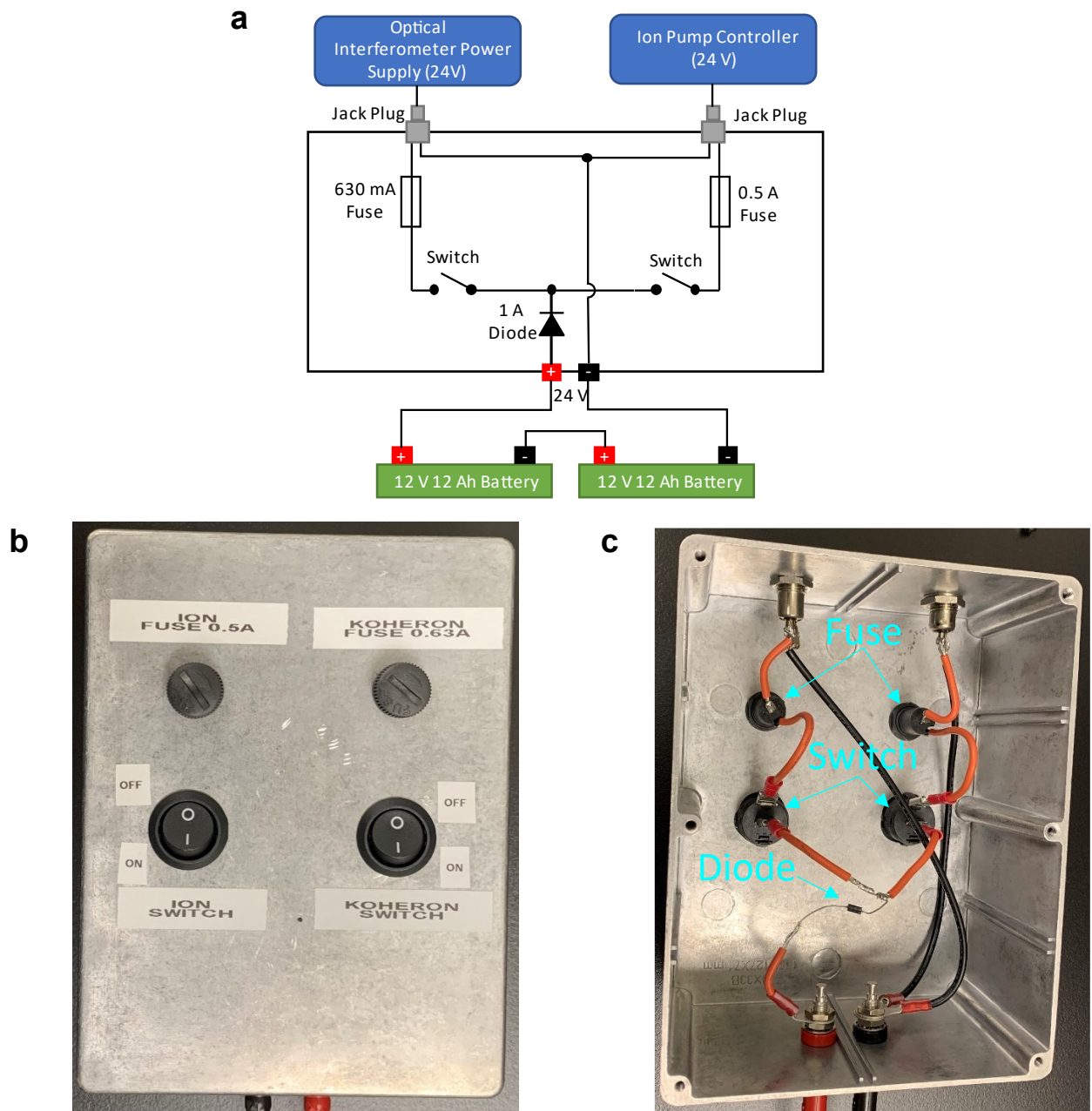

*Fig. S2. Power supply enclosure a) schematic, b) exterior, c) interior.*

# S9. EXPERIMENTAL CONDITIONS & RESULTS

All experiment conditions—i.e., date, time, location, cloudiness, relative humidity, vacuum chamber temperature ($T_{ch}$), ambient atmospheric temperature ($T_{atm}$), viewport field of view angle ($\theta_v$), and angle at which the sun is facing the membrane ($\theta_{sun}$)—and experimental results for the membrane average temperature shift ($\overline{\Delta T_m}$) are outlined in Table S3. The cloudiness level is determined visually by observing the sky directly above the experiment (see pictures in Fig. S3). The relative humidity is taken from Environment Canada's historical database for the weather and climate in Ottawa, ON [8]. Finally, $T_{ch}$ and $T_{atm}$ are determined using the thermistor inside the chamber and the temperature probe outside the chamber, respectively, $\theta_{sun}$ is determined as described in Supplementary Note 6, and $\theta_v$ is determined as described in Supplementary Note S9.1.

Table S3. Radiative cooling experiment conditions and experimental results.

| Test | Date | Time | Location[2] | Atmospheric Conditions | | $T_{atm}$ (K) | $T_{ch}$ (K)[3] | $\theta_v$ (rad) | $\theta_{sun}$ (rad) | $\overline{\Delta T_m}$ (K) |
|---|---|---|---|---|---|---|---|---|---|---|
| | | | | Cloudiness | Relative Humidity (%) | | | | | |
| 1[1] | 10-27-2021 | 21:15 | uOttawa | Cloudy | 72 | 284 | N/A | 0.45 | Night | -2.5 ± 0.8 |
| 2[1] | 10-28-2021 | 20:50 | uOttawa | Clear | 79 | 282 | N/A | 0.45 | Night | -4.2 ± 1.3 |
| 3 | 05-11-2022 | 22:10 | uOttawa | Cloudy | 41 | 298 | 298 | 0.68 | Night | -3.72 ± 1.2 |
| 4 | 05-11-2022 | 22:30 | uOttawa | Clear | 41 | 298 | 298 | 0.68 | Night | -5.5 ± 1.7 |
| 5 | 05-12-2022 | 9:30 | uOttawa | Clear | 35 | 298 | 297 | 0.68 | 0.72 | -4.7 ± 1.5 |
| 6 | 05-18-2022 | 11:30 | uOttawa | Clear | 41 | 294 | 290 | 0.68 | 0.46 | -3.9 ± 1.2 |
| 7 | 05-20-2022 | 11:20 | uOttawa | Cloudy | 65 | 293 | 295 | 0.68 | 0.45 | -1.8 ± 0.6 |
| 8 | 06-22-2022 | 10:25 | uOttawa | Clear | 75 | 300 | N/A | 0.68 | 0.54 | -5.1 ± 1.6 |
| 9 | 06-28-2022 | 23:20 | uOttawa | Clear | 77 | 295 | N/A | 0.68 | Night | -5.0 ± 1.6 |
| 10 | 08-02-2022 | 23:00 | uOttawa | Clear | 77 | 291 | N/A | 0.68 | Night | -6.5 ± 2.1 |
| 11 | 08-06-2022 | 10:20 | uOttawa | Clear | 73 | 303 | 300 | 0.68 | 0.60 | -4.6 ± 1.5 |
| 12 | 08-13-2022 | 13:45 | Rural-FNR | Clear | 43 | 300 | 300 | 0.68 | 0.66 | -9.0 ± 2.8 |
| 13 | 08-13-2022 | 16:30 | Rural-VKH | Clear | 38 | 302 | 303 | 0.68 | 1.1 | -9.3 ± 2.9 |
| 14 | 08-13-2022 | 22:25 | Rural-VKH | Clear | 63 | 289 | 292 | 0.68 | Night | -7.1 ± 2.2 |
| 15 | 08-15-2022 | 21:00 | uOttawa | Cloudy | 63 | 298 | 298 | 0.68 | Night | -2.5 ± 0.8 |

| 16 | 08-24-2022 | 11:20 | uOttawa | Clear | 74 | 301 | N/A | 0.68 | 0.62 | -4.0 ± 1.3 |
| 17 | 08-27-2022 | 13:00 | Rural-VKH | Slightly Cloudy | 53 | 297 | 295 | 0.68 | 0.66 | -4.3 ± 1.4 |
| 18 | 08-27-2022 | 13:20 | Rural-VKH | Slightly Cloudy | 53 | 297 | 298 | 0.68 | 0.66 | -6.5 ± 2.1 |
| 19 | 08-28-2022 | 00:10 | Rural-VKH | Clear | 90 | 289 | 291 | 0.68 | Night | -6.9 ± 2.2 |
| 20 | 08-28-2022 | 12:40 | Rural-VKH | Clear | 53 | 300 | N/A | 0.68 | 0.63 | -7.6 ± 2.4 |

1. Experiment 1 and 2 are the only ones where $\theta_v = 0.45$ because the membrane was mounted lower in the nipple of the vacuum chamber for these experiments.
2. FNR represents Fournier, ON, and VKH represents Vankleek Hill, ON.
3. N/A means the thermistor is not working so no $T_{ch}$ measurements were taken.

No picture was taken.     No picture was taken.     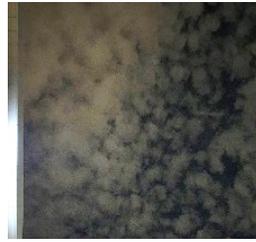 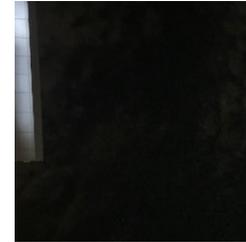 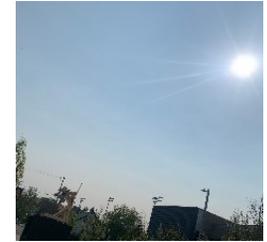

10-27-2021 @ 21:15     10-27-2021 @ 20:50     05-11-2021 @ 22:10     05-11-2021 @ 22:30     05-12-2022 @ 9:30

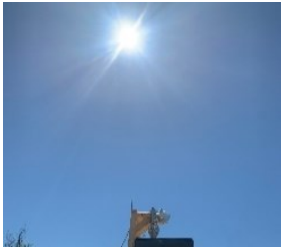 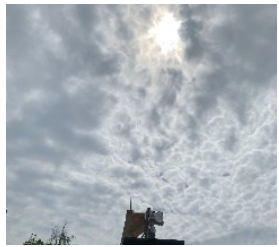 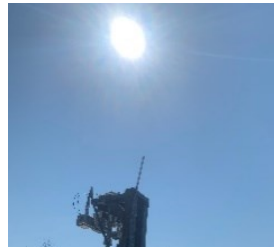 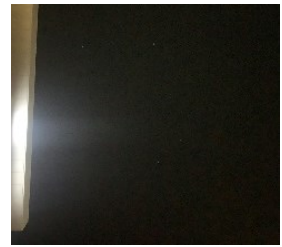 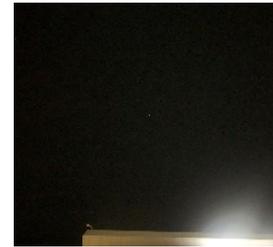

05-18-2022 @ 11:30     05-20-2022 @ 11:20     06-22-2022 @ 10:25     06-28-2022 @ 23:20     08-02-2022 @ 23:00

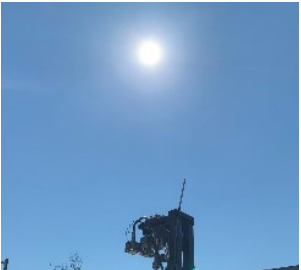 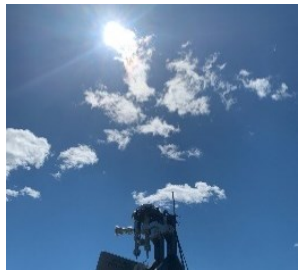 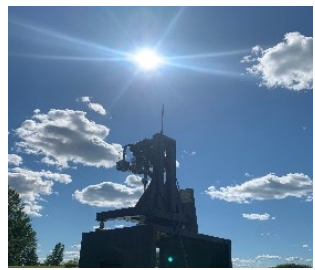 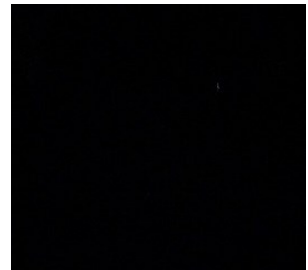 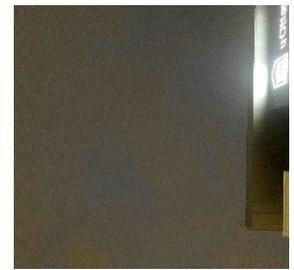

08-06-2022 @ 10:20     08-13-2022 @ 13:45     08-13-2022 @ 16:30     08-13-2022 @ 22:25     08-15-2022 @ 21:00

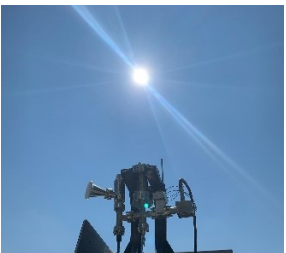 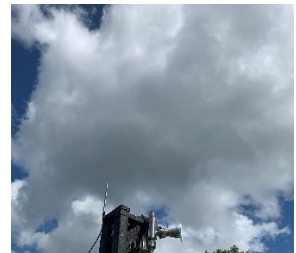 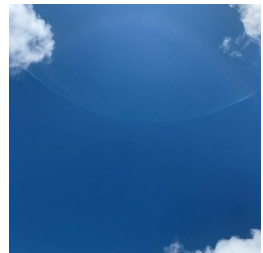 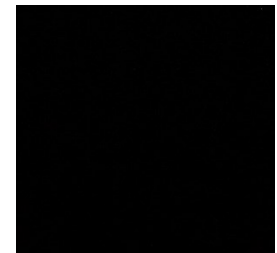 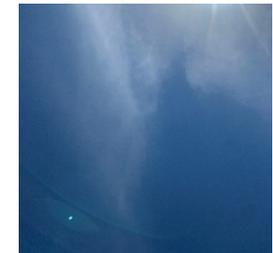

08-24-2022 @ 11:20     08-27-2022 @ 13:00     08-27-2022 @ 13:20     08-29-2022 @ 00:10     08-29-2022 @ 12:40

*Fig. S3. Pictures of the sky during each experiment. Some images have been adjusted for better contrast and brightness.*

## S9.1. Viewport Angle Measurements

The viewport angle ($\theta_v$) is calculated using simple trigonometry with CAD measurements of the mounting setup. From experiment 3 and onwards (refer to Table S3), the membrane mount location inside the nipple of the vacuum chamber is depicted in Fig. S4. Hence, the viewport angle is calculated as $\theta_v = \tan^{-1}\left(\frac{10.16}{12.59}\right) = 0.68$.

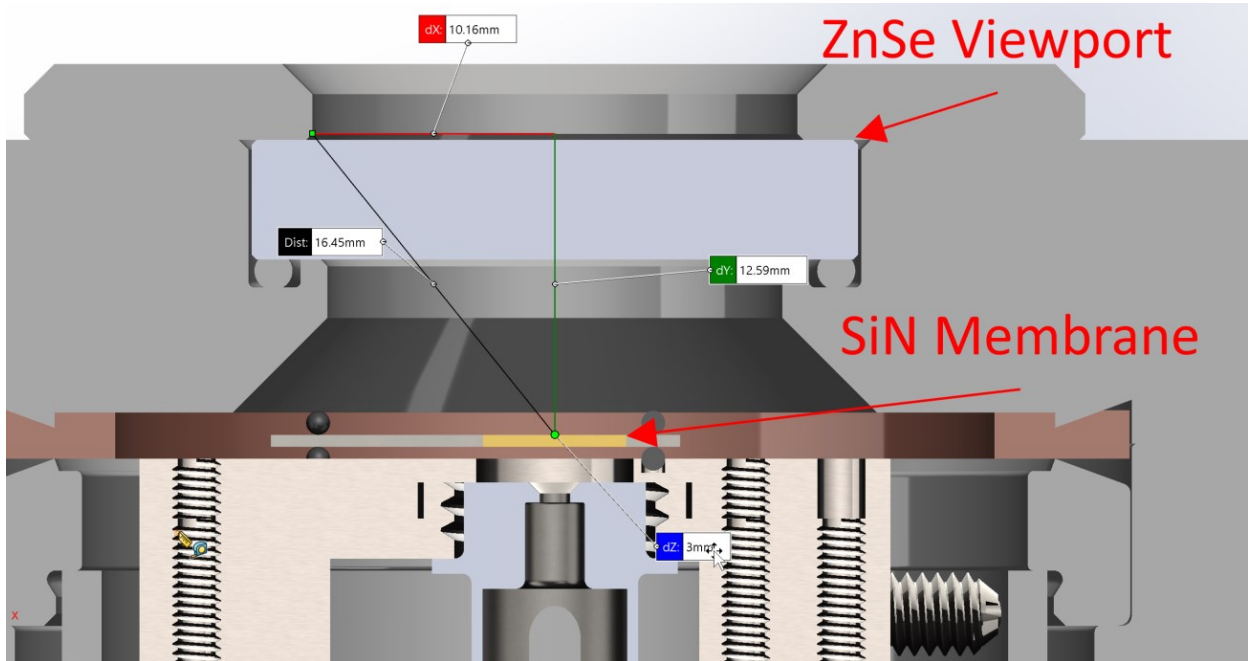

*Fig. S4. Viewport Angle CAD Measurements*

For experiments 1 and 2 (refer to Table S3), the membrane mount was located lower in the nipple of the vacuum chamber, with a dimension dY $\approx$ 21 mm. This means $\theta_v = \tan^{-1}\left(\frac{10.16}{21.00}\right) = 0.45$ for these experiments.

## S10.   SILICON FRAME TEMPERATURE UNDER DIRECT SUNLIGHT

Fig. S5 shows that the silicon frame can reach significantly higher temperatures than the ambient environment when exposed to direct sunlight due to its high absorptivity in the solar spectrum.

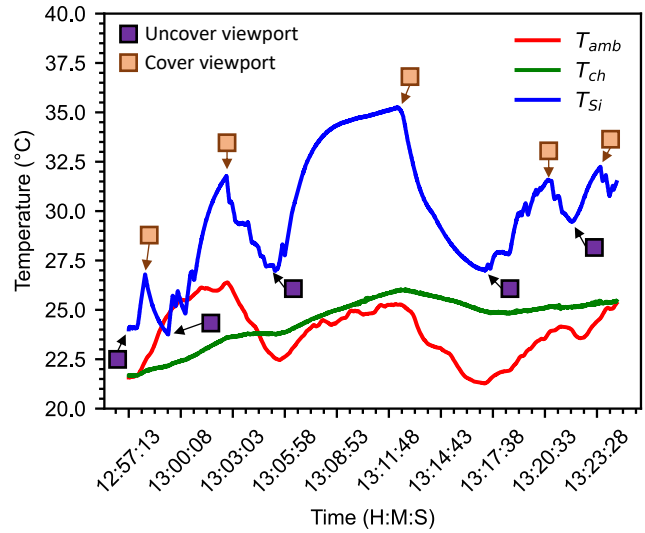

*Fig. S5. Ambient ($T_{amb}$), chamber ($T_{ch}$), and silicon frame ($T_{Si}$) temperature during experiment conducted on 2022-08-27 under direct sunlight.*

# SUPPLEMENTARY INFORMATION REFERENCES